\documentclass[runningheads]{llncs}
\usepackage{color}
\usepackage{hyperref}
\usepackage{amsmath,amsfonts,amssymb}
\usepackage{graphicx}
\usepackage{wrapfig}
\begin{document}
\title{fMRI-Kernel Regression: A Kernel-based Method for Pointwise Statistical Analysis of rs-fMRI for Population Studies\thanks{This work is supported by the following grants: R01-NS074980, W81XWH-18-1-0614, R01-NS089212, and R01-EB026299.}}

\author{Anand A. Joshi \and 
Soyoung Choi \and
Haleh Akrami \and 
Richard M. Leahy}

% index{Joshi, Anand A.}
% index{Akrami, Haleh}
% index{Li, Jian}
% index{Leahy, Richard M.}

\institute{University of Southern California, Los Angeles, CA, USA}

\authorrunning{fMRI-Kernel Regression}
\maketitle              % typeset the header of the contribution
\begin{abstract}
Due to the spontaneous nature of resting-state fMRI (rs-fMRI) signals, cross-subject comparison and therefore, group studies of rs-fMRI are challenging. Most existing group comparison methods use  features extracted from the fMRI time series, such as connectivity features, independent component analysis (ICA), and functional connectivity density (FCD) methods. A recently developed method based on Brainsync transform allows the group synchronization of data and direct comparison of rs-fMRI time series. However, in group studies, especially in the case of spectrum disorders, distances to a single atlas or a representative subject do not fully reflect the differences between subjects that may lie on a multi-dimensional spectrum. Moreover, there may not exist an individual subject or even an average atlas in such cases that is representative of all subjects. Here we describe an approach that measures pairwise distances between the synchronized rs-fMRI signals of pairs of subjects instead of to a single reference point. We also present a method for fMRI data comparison that leverages this generated pairwise feature to establish a radial basis function kernel matrix. This kernel matrix is used in turn to perform kernel regression of rs-fMRI to a clinical variable such as a cognitive or neurophysiological performance score of interest.  This method opens a new pointwise analysis paradigm for fMRI data. We demonstrate the application of this method by performing a pointwise analysis on the cortical surface using rs-fMRI data to identify cortical regions associated with variability in ADHD index. 
 While pointwise analysis methods are common in anatomical studies such as cortical thickness analysis and voxel- and tensor-based morphometry and its variants, such a method is lacking for rs-fMRI and could improve the utility of rs-fMRI for group studies. The method presented in this paper is aimed at filling this gap.

\keywords{rs-fMRI \and statistical methods \and kernel-based methods}
\end{abstract}
\section{Introduction}

The brain consumes 60 to 80\% of its energy in resting periods during which it produces low-frequency spontaneous modulations of the BOLD signal measured using fMRI \cite{smitha2017resting,van_den_heuvel_exploring_2010}. The fMRI signal acquired during rest (rs-fMRI) has been used extensively to measure functional connectivity between different brain regions as reflected in the correlation between rs-fMRI signals at difference locations. \cite{horwitz_elusive_2003,lang_resting-state_2014,smith_network_2011,smitha_resting_2017,van_den_heuvel_exploring_2010}. Additionally, the rs-fMRI signal provides a rich marker for underlying cytoarchitecture and functional connectivity of the brain and has been used for brain parcellation \cite{chong2017individual,glasser2016multi}. These properties make rs-fMRI signal useful for identifying functional differences in groups as well as for finding functional correlates of clinical measurements such as performance scores on cognitive and neuropsychological studies. Functional connectivity measured from rs-fMRI also allows longitudinal studies of brain development and can serve as a diagnostic biomarker in cross-sectional studies of various neurological and psychological diseases and conditions \cite{redcay_intrinsic_2013}. 

Since rs-fMRI data reflect spontaneous brain activity, it is not possible to directly compare resting-state signals across subjects due to a lack of temporal synchronization. Instead, comparisons typically make use of connectivity features \cite{iraji_connectivity_2016}, typically computed from pairwise correlations of the rfMRI time-series between a point of interest and other locations in the brain \cite{fan_human_2016}. A majority of fMRI studies have focused on group differences in functional connectivity \cite{tomasi2010functional,iraji2016connectivity}, independent component analysis (ICA) \cite{calhoun2001method,erhardt2011comparison},  brain networks and roi-wise analysis.  Relatively few group studies have studied pointwise differences in the brain across groups \cite{tomasi2010functional,thirion2007analysis,beckmann2009group}. Conversely, pointwise analysis methods are common in anatomical studies such as cortical thickness analysis, voxel- and tensor-based morphometry and its variants. Similar pointwise analysis in  rs-fMRI is uncommon, but can greatly expand the utility of rs-fMRI for group studies \cite{akrami_group-wise_2019,lorbert2012kernel,haxby2011common}. 

The Brainsync transform \cite{joshi_are_2018} is an orthogonal transform that represents fMRI data on a hypersphere, and performs a temporal alignment of time-series at homologous locations between pairs or groups of subjects to synchronize their rs-fMRI data. As a result of synchronization the time series are approximately equal at homologous locations across the group. This transform allows us to compare subjects time-series on a pointwise basis. Here we present an fMRI-kernel regression method that builds on this ability to perform point-wise analysis. Specifically, we use a metric based on pairwise subject differences of rs-fMRI data in conjunction with a kernel to perform regression to a clinical variable at each point on the cortex \cite{von_gadow_regression_2001}. 

We demonstrate the approach in an application to identifying brain locations that are associated with an Attention-Deficit/Hyperactivity Disorder (ADHD) index. ADHD is a neurodevelopmental disorder characterized by the presence of high levels of hyperactive, impulsive, and inattentive behaviors and  is most common in children, but often persists into adolescence and adulthood \cite{c}. Several individual studies have been able to detect low brain activity in the frontal lobe in subjects with ADHD in comparison to controls \cite{mccarthy2014identifying}. In addition, alterations of the default mode and cognitive control networks have been observed \cite{sutcubasi2020resting}. However, in attempting to identify consistent observations of neural dysfunction in ADHD subjects, meta-analysis revealed high variance in study outcomes \cite{sutcubasi2020resting,mccarthy2014identifying}. Here we identify functionally altered cortical regions in ADHD subjects compared to the control subjects using two approaches.
We first perform a pairwise analysis between 2000 random pairs of subjects in which the differences in their ADHD indices is correlated against the distance between their synchronized time-series. We compare this to nonlinear fMRI-kernel regression using a kernel matrix that is generated from the pairwise distances between fMRI signals. 

\section{Materials and Methods}
First, we briefly review the Brainsync transform and then present the two approaches for pointwise fMRI analysis: correlation of pairwise distances and kernel regression.

\subsection{Data and Preprocessing}
\label{sec. data_preprocessing}
The rfMRI data was collected as a part of the ADHD-200 Global Competition (Peking University data) \cite{cao_abnormal_2006} and preprocessed using the BrainSuite Functional Pipeline (\url{brainsuite.org}). A group of 200 subjects with ADHD measured using the ADHD Rating Scale-IV (ADHD-RS) was selected. We used 150 test subjects comprised of 85 ADHD subjects (age=$12.0 \pm 2.0$; 75M:10F; ADHD Index=$50.6\pm 8.5$) and 65 control subjects (age=$11.1 \pm 1.8$; 39M:26F; ADHD Index=$29.2\pm 6.3$) to test the relationship of ADHD Index with synchronized rfMRI using two methods (pairwise regression and kernel regression). Images were acquired using
a Siemens Trio 3-Tesla scanner. All of the resting-state functional data were acquired using an echo-planar imaging (EPI) sequence as described in Cao et al. (2006) \cite{cao_abnormal_2006}.
We used the BrainSuite fMRI Pipeline (BFP) to process the rfMRI subject data and generated grayordinate representations of the preprocessed rfMRI signals \cite{glasser2013minimal}. BFP is a software workflow that processes fMRI and T1-weighted MR data using a combination of software that includes BrainSuite, 
AFNI, FSL, and MATLAB scripts to produce processed fMRI data
represented in a common grayordinate system that contains both cortical surface vertices and subcortical volume voxels. Starting from raw T1 and fMRI images, BFP produces processed fMRI data
represented both on surface and volume co-registered with BrainSuite's BCI-DNI atlas as well as a grayordinate based representation.

\subsection{The Brainsync Transform}
\begin{figure}
    \centering
    \includegraphics[width=1\textwidth]{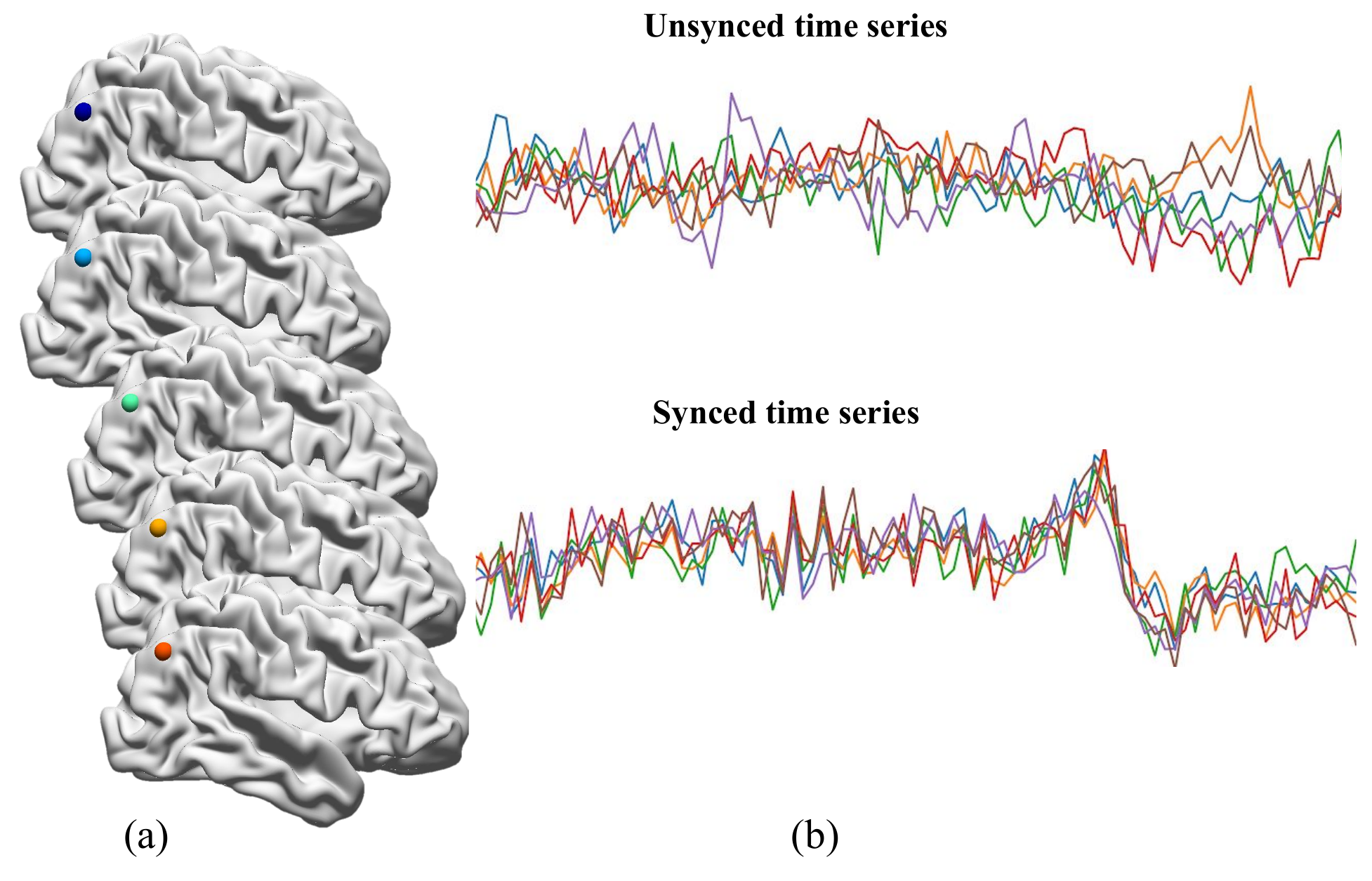}
    \caption{(a) Homologous locations on a group of subjects; (b) (top) rs-fMRI time-series at those locations; (bottom) the same time series after application of the Brainsync transform}
    \label{fig:brainsync}
\end{figure}

Unlike cortical thickness, which is popularly used for morphometric studies, the 
rs-fMRI signal is a multivariate descriptor representing spontaneous brain activation, and is not directly comparable across subjects.  Brainsync finds an orthogonal transform between  time-series data for two or more subjects to find an optimal temporal alignment between anatomically homologous points in the brain \cite{joshi_are_2018}. Fig. \ref{fig:brainsync}
 shows an example of the synchronization process. The time-series data from multiple subjects at a single homologous point on the cortex is labeled by subject with different colors. These time series are synchronized to that of an individual subject as an illustrative example. This in turn allows direct comparison of functional data at corresponding points in the brain across subjects. 
 
As input, we assume that the rs-fMRI data is available in the grayordinate system defined by the Human Connectome Project (HCP). This is a standardized representation of gray matter in the brain as a combination of a tessellated cortical surface and a voxel-wise map of subcortical gray matter \cite{glasser2013minimal}. The fMRI data are assumed aligned to a common grayordinate representation, which for each subject consists of a matrix $F$ of size $T \times V$, where $V$ is the number of vertices in the cortical mesh plus the number of subcortical voxels and $T$ is the number of time points. Hence the corresponding columns in $F$ represent the time-series at homologous locations in the brains. The data vectors in each column are normalized to have zero mean and unit norm.

Given data from two subjects denoted by $X_{T\times V}$ and $Y_{T\times V}$ respectively, Brainsync finds an orthogonal matrix $O_{XY}$ such that $\|X-OY \|_F^2$ is minimized where $\|\cdot \|_F$ is the Frobenius norm. The data $X$ and synced data $OY$ are comparable pointwise (at each element $v \in V$). $O_{XY}$ can be computed in closed-form from the SVDs of $X_{T\times V}$ and $Y_{T\times V}$.

%Next we extend this ability to compare pairs of scans pointwise for a regression study.

\subsection{Pairwise Distance Correlation}
\label{sec:pairwise_regression}
First, we present a naive pairwise correlation method, then extend this method using kernel regression.
We performed a pairwise correlation analysis by synchronizing then computing distance measures between 2000 random pairs from the 150 subjects. Let's represent the $T\times V$ fMRI data for the $i^{th}$ subject at vert $v$ and time $t$ by $F_i(t,v)$ and its corresponding ADHD index by $y_i$. The distance between the synchronized pairs of subjects $i$ and $j$ is used as a statistic at point $v$ $d_F(i,j) = \sum_t (F_i(t,v)-O_{ij} F_j(t,v))^2$. We also computed the difference between the ADHD indices of the two subjects $d_T(i,j) = |y_i - y_j|$. We then computed Pearson correlations between $d_F$ and $d_T$ and converted to $p$-values using a permutation test (2000 permutations). Benjamini-Hochberg FDR correction was performed to correct for multiple comparisons across vertices.

%%%To test the reproducibility of the tests, we repeated the entire experiment for the pairwise statistic, again with 2000 randomly selected pairs. For additional verification that the significance that we find is not by chance, randomly permuted ADHD indices were assigned to the subjects and the test repeated.

\subsubsection{Limitations of the pairwise correlation approach}
\begin{wrapfigure}{l}{0.5\textwidth}
    \includegraphics[width=0.50\textwidth]{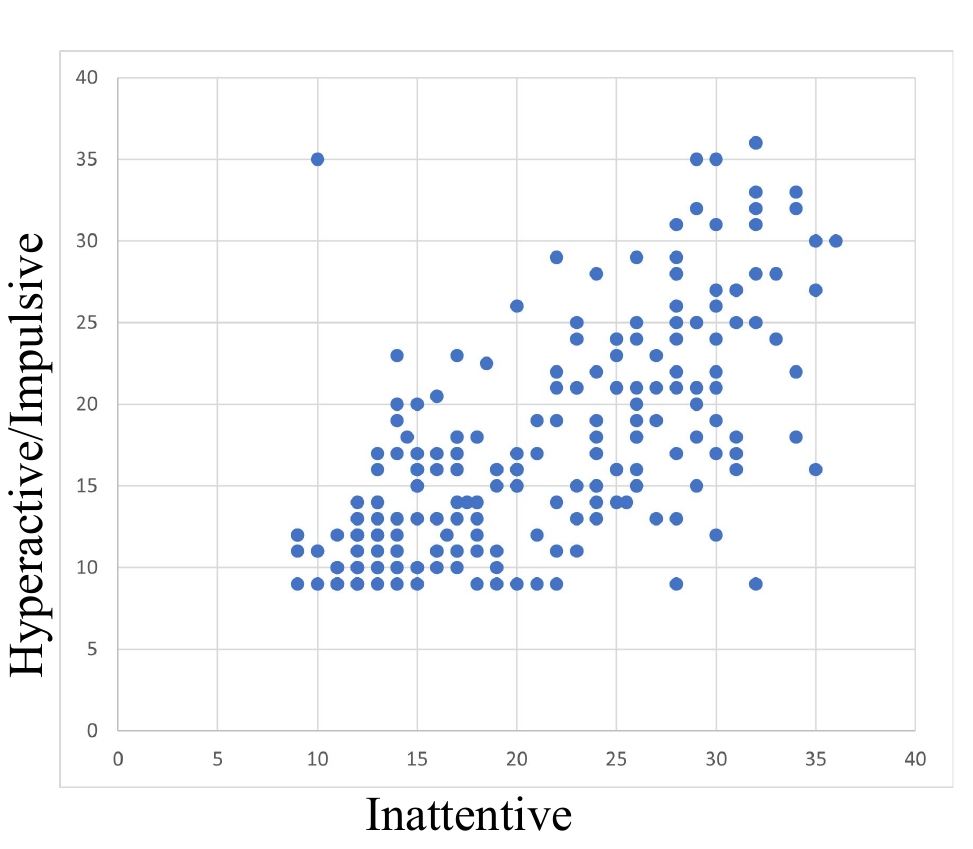}
    \caption{A scatter plot of the two components of the ADHD index: hyperactive/impulsive index and inattentive index.}
    \label{fig:adhd_index spread}
\end{wrapfigure}
In complex spectrum disorders such as ADHD, the neuropsychological index  may not map to a simple 1D subspace. The ADHD index is a summation of two indices:  hyperactivity/impulsivity and inattentiveness, for which the scatter plot is shown in Fig. \ref{fig:adhd_index spread}. It is clear from this plot that two subjects with the same ADHD index can differ in their underlying neural states: while the measures are correlated there can be a substantial deviation between the two measures for each individual. More generally, we can postulate that the distribution of ADHD indices lies on a multidimensional manifold. To capture the characteristics of this manifold we can use pairwise distances between subjects and a kernel-based approach to model a nonlinear relationship between ADHD index and rs-fMRI activity.

\subsection{fMRI-Kernel Regression}
\label{sec:fMRI-kernel_regression}
The pairwise distances between fMRI signal at each location obtained using the Brainsync transform can be used to generate the kernel matrix to fully account for the signal space.
%In the previous analysis we compared the pairwise distance in fMRI signal to the distance in clinical variables. However, as noted, this does not fully account for the signal space. Even if the rs-fMRI signal is not directly comparable across subjects, the distance between them can be computed using the brainsync transform. 
\begin{figure}[t]
    \centering
    \includegraphics[width=1\textwidth]{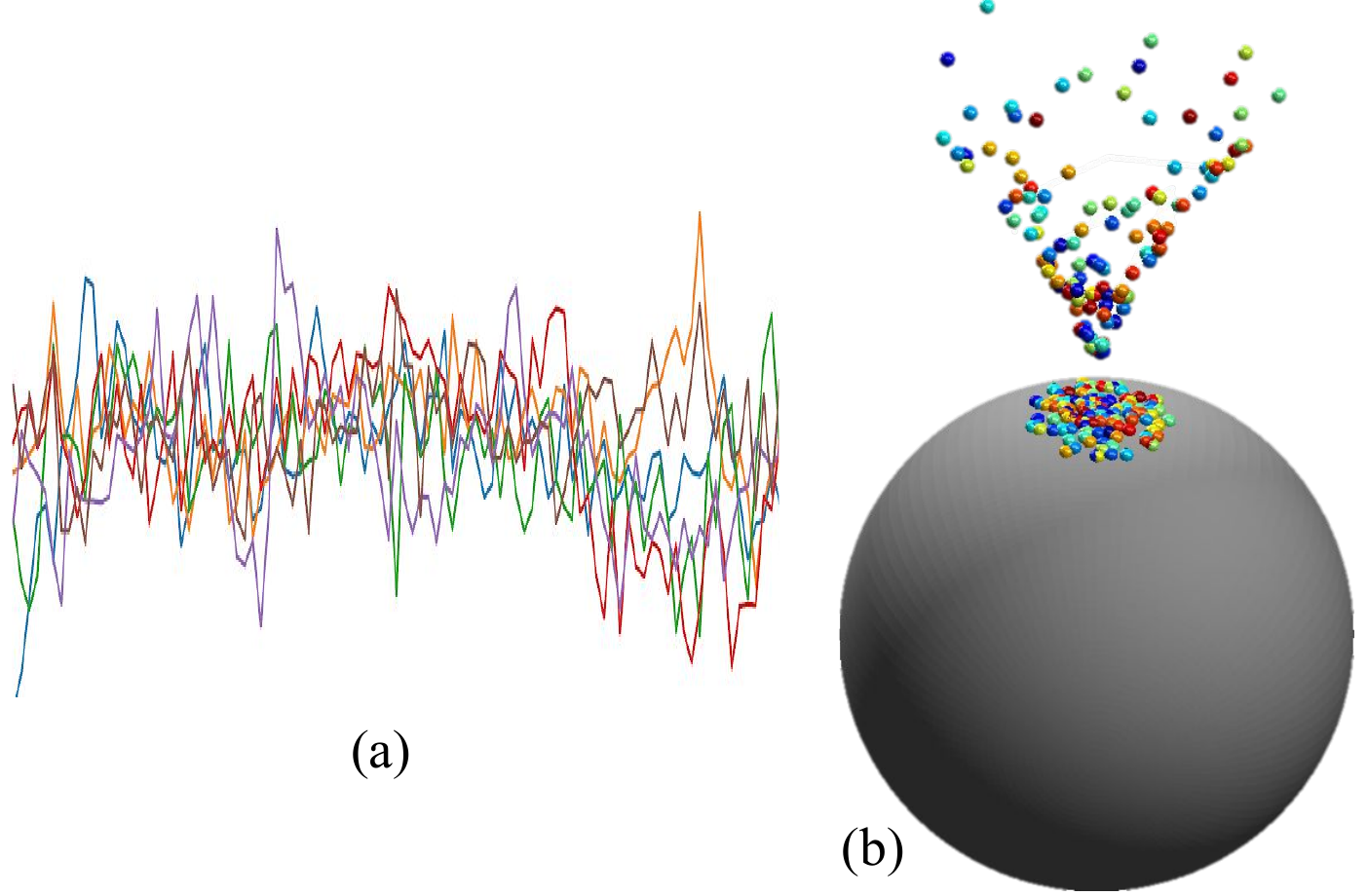}
    \caption{(a) the rs-fMRI time series data from multiple subjects at a given point on the cortex is represented with different colors. (b) The Brainsync transform embeds this data on a hypersphere, represented as colored points on the surface of a 3D sphere for illustration. Distances between pairs of points on the sphere measures similarity between rs-fMRI signal. The application of the kernel transforms the data to a higher dimension represented by points above the sphere.}
    \label{fig:kernel_lift}
\end{figure}
The fMRI signal can be represented on a sphere \cite{joshi_are_2018} and therefore distance between them can be computed on the sphere, instead of in the ambient Euclidean space. We define a distance $d(F_i,F_j)$ between two rs-fMRI timeseries, represented as vectors, $\vec{F}_i(v) = F_i(\cdot,v)$ and $\vec{F}_j(v) = F_j(\cdot,v)$ at point $v$ on the cortex as:
\begin{align}
d(F_i,F_j)(v) = \arccos (\vec{F}_i^T(v) O_{ij} \vec{F}_j(v))   
\end{align}
where $O_{ij}$ representes the $T\times T$ orthogonal matrix found by the Brainsync transform that synchronizes the two rs-fMRI datasets.

We use an extension of the Nadaraya-Watson nonlinear kernel regressor \cite{hardle1990applied} to the hypersphere, with the kernel function to generate kernel matrix at point $v$ on the cortex as
\begin{align}
K(v) = \exp{(\gamma d(F_i,F_j)(v))},    
\label{eq:kernel matrix}
\end{align} 
where $\gamma$ is the bandwidth parameter. Interestingly, the kernel PDF represented by this kernel matrix is equivalent to the wrapped Gaussian PDF on the hypersphere associated with square-root histograms, given a cohort comprising the independent variables. This kernel lift to higher dimension as depicted in Fig. \ref{fig:kernel_lift}.

At each location $v$ on the cortex, we want to test the hypothesis that the clinical variable is dependent on the fMRI signal. For this purpose, we use an extension of the Nadaraya-Watson kernel regression \cite{hardle1990applied} to the space of rs-fMRI, with the kernel PDF given by $PK(F,\mu,\gamma) = \frac{1}{eta}\exp( -\gamma d(F,\mu))$, where $\mu$ is the normalization constant, $\gamma$ is the concentration parameter and $\eta$ is the normalization constant. This corresponds to the kernel matrix given in Eq. \ref{eq:kernel matrix}. Given a population comprising the independent variables as the rs-fMRI data $\{F_i\}_i=1^N$ of $N$ subjects, and the clinical variables (ADHD index in our case) as the dependent variable $\{y_i\}_i=1^N$, the clinical variable $y$ for an arbitrary rs-fMRI data $F$ is given by 
\begin{align}
    \hat{y} = r(F,{F_i,y_i}_{i=1}^N) = \frac{\sum_{i=1}^N PK(F;F_i,\gamma)y_i}{\sum_{i=1}^N PK(F;F_i,\gamma)}.
\end{align}
We term this procedure \emph{fMRI-Kernel regression}.
\subsubsection{Choice of Kernel Bandwidth}
The choice of kernel bandwidth $\gamma$ is critical  for reliable regression. We use a leave-one-out (LOO) cross validation approach \cite{hardle1990applied} to automatically optimize the bandwidth parameter We optimize $\gamma$ by grid search; we first compute the mean squared error, in the LOO regression estimates across the dataset, for a sequence of discretized parameter values and then select the parameter value that minimizes this error \cite{hardle1990applied,hardle1994applied}. In our case this process resulted in $\gamma = 2.6$. This choice was consistent for different sizes of populations as well as different cortical vertices.

\subsection{Statistical Hypothesis Testing} 
The pointwise parametric hypothesis testing in the framework of general linear models runs a test at each voxel and adjusts p-values to control for Type-I error arising from multiple comparisons, using Gaussian field theory. However, these parametric approaches make strong assumptions on the data distributions and the local dependencies. In contrast, permutation tests are nonparametric, rely on the assumption of exchangeability, and are more robust to deviations of the data and effects of processing from an assumed model \cite{oden1975arguments}. For this reason, we use permutations for hypothesis testing.
We use a permutation test to test the null hypothesis that there is no relationship between the rs-fMRI data $\{F_i\}_{i=1}^N$ and the clinical variable $\{y_i\}_{i=1}^N$.
For this purpose, we compare the residuals of the pairwise regression (Sec. \ref{sec:pairwise_regression}) or Kernel regression (Sec. \ref{sec:fMRI-kernel_regression}) obtained on the population data with the residuals obtained using the permuted clinical variables and the same regression procedures, respectively.
If the null hypothesis was true, i.e., the clinical variable was independent of the fMRI signal, then the residuals obtained using the permuted clinical variables would have the same variance as the residuals obtained using the unpermuted clinical variables. Thus, we propose the test statistic as the well-known F-statistic, i.e., the ratio of the sample variances of the two sets of residuals.
We compared the variance in the case of permuted and unpermuted clinical variables using an $F-test$. This resulted in  $p$-values at each point on the cortex. The values were corrected for multiple comparisons using the Benjamini-Hochberg procedure \cite{benjamini1995controlling}.

\subsection{Simulation Study}
\begin{figure}
    \centering
    \includegraphics[width=1\textwidth]{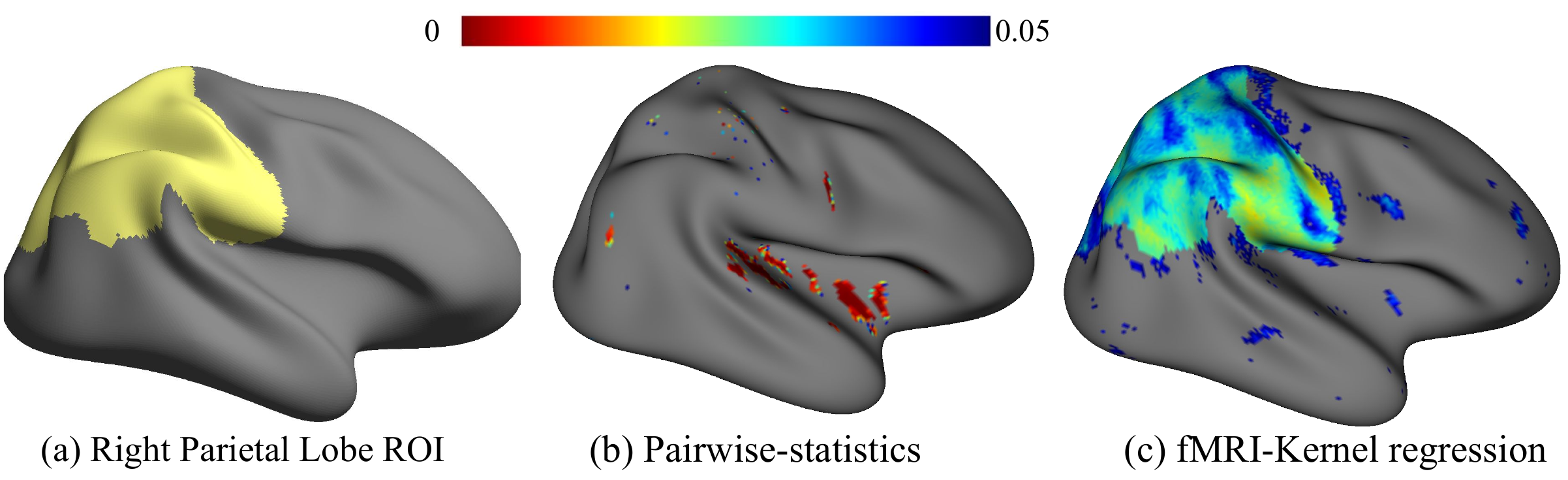}
    \caption{A simulation showing advantage of fMRI kernel regression over pairwise statistics approach: (a) we selected a ROI (right parietal lobe) and add  iid Gaussian noise ($\sigma = 0.3$); (b) results of pairwise statistics fail to show the significance in the ROI; (c) the fMRI-Kernel regression shows significance in the ROI where the noise was added.}
    \label{fig:simulation}
\end{figure}

To demonstrate the intuition and the motivation behind the fMRI-kernel regression approach, we describe a simulation in this section where we modify the rs-fMRI signal in a known manner in a specific brain region. While this demonstration is not meant to be realistic, it demonstrates the utility of kernel regression when multivariate descriptor such as a rs-fMRI signal is regressed against a univariate clinical variable. 
We consider the population of 50 subjects under consideration, and randomly shuffle their ADHD indices to avoid bias.  
Then, we selected a region in the brain, specifically vertices on the cortex corresponding to the parietal lobe. For each of the subjects, we added a 0 mean Gaussian noise with standard deviation proportional to the normalized ADHD index, to the rs-fMRI signal in each vertex in the parietal lobe.
Then we performed pairwise distance correlation (Sec. \ref{sec:pairwise_regression}) as well as fMRI-kernel regression (Sec. \ref{sec:fMRI-kernel_regression}) approaches to the data. The results of the analysis are shown in Fig. \ref{fig:simulation}. It can be seen in Fig/ \ref{fig:simulation}(b,c) that the pairwise statistical analysis failed to identify the region where the noise was added to the data, whereas the fMRI-kernel regression was able to correctly identify that region. This is because in the case of multivariate descriptors such as rs-fMRI, adding noise results in an additional random angle in the descriptor. The pairwise descriptor fails to account for the angle in the multivariate descriptor whereas the kernel-based analysis does not have that limitation since it works in a higher dimension 
In the case of ADHD, this property is useful since, two ADHD subjects with same ADHD index might have a completely different reason for their condition, and their brain correlates might be different, even for the same brain region.

\subsection{ADHD Study}
Finally, we applied the pairwise correlation (Sec. \ref{sec:pairwise_regression}) and the fMRI-kernel regression (Sec. \ref{sec:fMRI-kernel_regression}) to the ADHD dataset. We did this analysis for a larger  population of $N=150$ subjects and a smaller population of $N=50$ subjects.
For the larger population, the results of the 
 statistical tests showed large, highly significant clusters across the frontal, temporal and insular cortices, even after FDR correction. Significant, yet sparser clusters of regions were also found posteriorly. The resulting spatial map of the cortex shows an association of executive function networks to the ADHD indices. Similar results were also found for the fMRI-kernel regression tests. However, for a smaller population of $N=50$ subjects, the pairwise test shows a much noisier result, whereas the fMRI-kernel regression shows similar significant regions. 
 
\begin{figure}[t]
    \centering
    \includegraphics[width=1\textwidth]{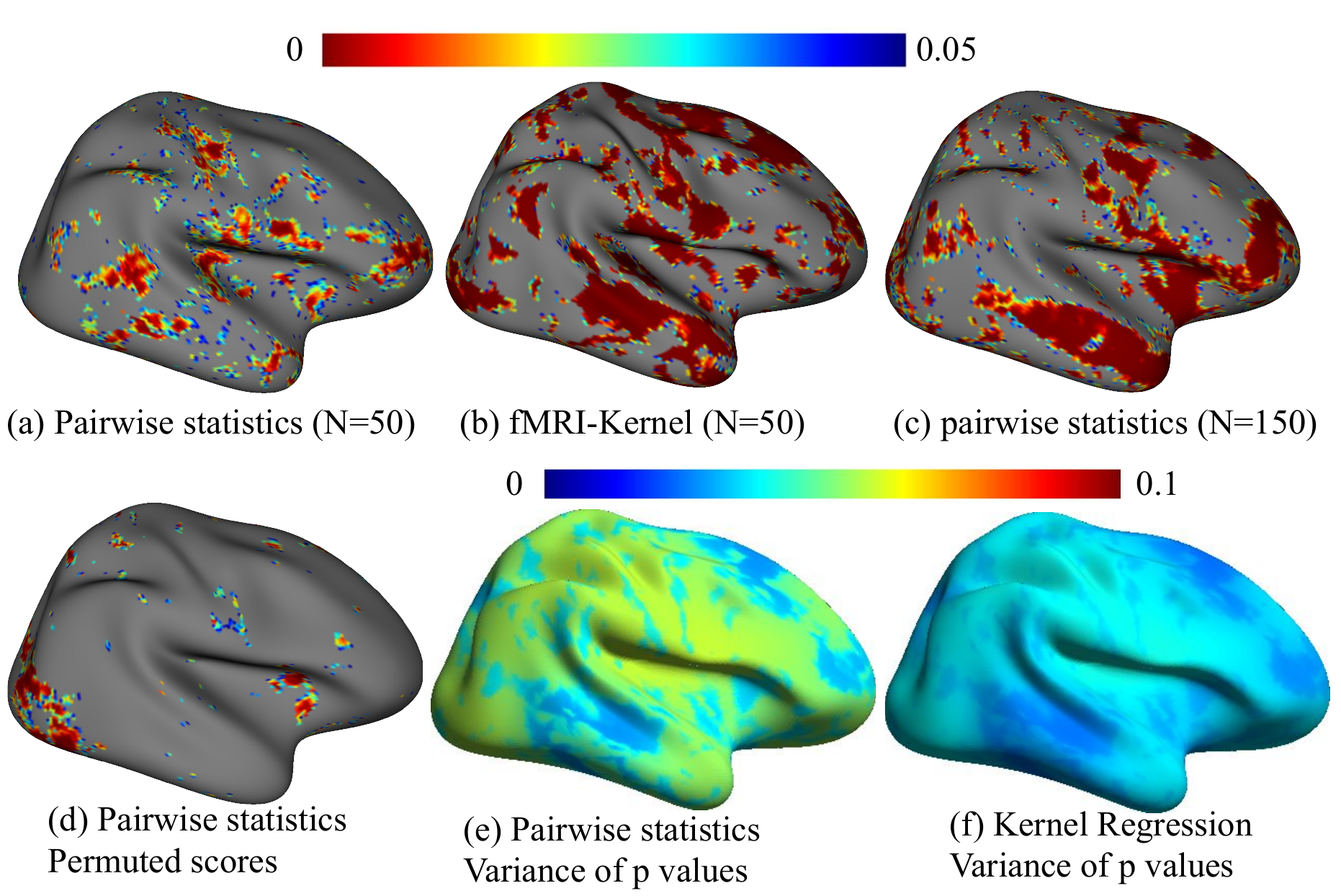}
    \caption{The results of the ADHD study: (a) pairwise statistical analysis with 50 subjects; (b) fMRI-kernel regression with 50 subjects; and (c) pairwise statistical analysis with 150 subjects; It can be seen that for a smaller population, the kernel-based method is able to find  the ADHD associated regions that are consistent with the results of 150 subject pairwise study. (d) The results of the pairwise test when ADHD index was randomly permuted for subjects still shows some false-positive regions. There were no false-positive regions for the kernel-based method. (e,f) The tests were repeated 10 times to get a variance of p-values. the kernel-based method shows lower variance of p-values indicating stability of the results.}
    \label{fig:adhd_results}
\end{figure}

For additional confirmation, the ADHD index was permuted and the test was repeated for the population of N=50. The pairwise statistical test showed some false-positive regions (Fig. \ref{fig:adhd_results}(d)) whereas no false-positive regions were found using the kernel-based method.

To evaluate the stability of the p values under variation in cohorts, we compute a set of p values by bootstrap sampling the original cohort (nboot=10) and analyzing the associated variation. The bootstraping procedure showed reduced variance of p-values in the case of fMRI-kernel regression (Fig. \ref{fig:adhd_results} (c,d)) whereas the smaller population size results demonstrate the utility of fMRI-Kernel regression for population studies of fMRI where $N$ is relatively small (Fig. \ref{fig:adhd_results}(a-c)). 

\section{Discussion}
\label{sec:discussion}
The rs-fMRI data provides a high dimensional feature at each point in the brain, and group comparisons of these features is possible after synchronization. For comparing distributions in high dimensional spaces, use of reproducing kernel Hilbert space (RKHS) has been described in the literature \cite{sejdinovic_hypothesis_2012,szekely_testing_2004,hall2002permutation}. Our current analysis on synchronized rs-fMRI shows that using the RKHS leads to further gains in statistical power. 

Another possible framework for analyzing the spherical representation generated by the Brainsync transform is to use directional statistics \cite{mardia2009directional}. There are parametric approaches such as Fisher-Bingham PDFs on $\mathbb{S}^n$ that are able to model generic anisotropic distributions, but might be intractable. The alternative is to use the von Mises-Fisher distribution, but it can only model istotropic distributions on the sphere \cite{mardia1976bayesian}. Awate et al. 2017  \cite{awate2017kernel} proposed an approximation of the normal law and modeling anisotropy through a covariance parameter on a Riemannian space. This approach is readily applicable to fMRI through the use of the Brainsync transform proposed here. However, for this purpose, a group synchronization of fMRI data as proposed in Akrami et al. \cite{akrami_group-wise_2019} would be necessary.
In this paper, we focused on kernel-based methods since they allow the use of pairwise distance computation using the kernel trick directly.

\section{Conclusion}
\label{sec:conclusion}
We presented a kernel-based approach and a novel metric for pointwise group comparison for rs-fMRI. Similar to anatomical analysis methods such as VBM/TBM and cortical thickness analysis, this method allows pointwise analysis of population rs-fMRI data.
We demonstrated the utility of the method for identifying cortical brain regions associated with ADHD. The fMRI-kernel regression shows improved performance over a naive pairwise analysis method by showing decreased bootstrap variance of p-values indicating the stability of the results.
 
\bibliographystyle{splncs04}
\bibliography{references}
\end{document}